\documentclass[10pt]{article} 
\usepackage[utf8]{inputenc}
\usepackage{opex3}
\usepackage[OT2,T1]{fontenc,rotating}
 \usepackage[english]{babel}
\usepackage{graphicx}

\usepackage{xspace,amsmath,amssymb}
\usepackage{textcomp}

\newcommand{\unit}[1]{\,\text{#1}\xspace}

\newcommand{\mW}{\unit{{mW}}}

\newcommand{\mm}{\unit{mm}}
\newcommand{\nm}{\unit{nm}}
\newcommand{\rad}{\unit{rad}}

\newcommand{\mum}{\,\ensuremath{\text{\textmu{}m}}\xspace}

\DeclareSymbolFont{cyrletters}{OT2}{wncyr}{m}{n}
\DeclareMathSymbol{\Sha}{\mathalpha}{cyrletters}{"58}

\newcommand{\requestreviewer}[1]{{#1}}

\newcommand{\ie}{{i.e.},\xspace}
\newcommand{\MAP}{\textsc{map}\xspace}
\newcommand{\modsqhada}[1]{| {#1} |^2}
\newcommand{\hadamard}{\odot}
\newcommand{\complex}{\mathbb{C}}
\newcommand{\derpar}[2]{\frac{\partial {#1}}{\partial {#2}}}
\newcommand{\ert}{\ensuremath{\varepsilon^2}\xspace}

 \newcommand{\CAMELOT}{\textsc{camelot}\xspace}
 
\newcommand{\PD}{\textsc{PD}\xspace}
\newcommand{\nv}[1]{\text{\boldmath$#1$}\xspace}
\newcommand{\DFT}[1]{\textsc{DFT}[#1]}
\def\Brack#1{\left[{#1}\right]} 
 \def\Paren#1{\left({#1}\right)}
\newcommand{\modc}[1]{| {#1} |}
\newcommand{\modcsq}[1]{| {#1} |^2}

\newcommand{\etal}{\emph{et al.}\xspace}
\newcommand{\IDFT}[1]{\textsc{IDFT}[#1]}
\renewcommand{\PD}{phase diversity\xspace}

\graphicspath{{./}}

\begin{document}

\title{Laser beam complex amplitude measurement by phase
  diversity}

\author{Nicolas V\'{e}drenne$^*$, Laurent M. Mugnier, Vincent Michau,
  Marie-Th\'er\`ese Velluet and Rudolph Bierent}

\address{Office National d'Etudes et de
  Recherches A\'{e}rospatiales (ONERA) \\ High Angular Resolution Unit, Optics Department,\\ 29 avenue
  de la division Leclerc\\
92332 Ch\^{a}tillon, FRANCE}

\email{$^*$nicolas.vedrenne@onera.fr} 



\begin{abstract}
  The control of the optical quality of a laser beam requires a complex
  amplitude measurement able to deal with strong modulus variations and
  potentially highly perturbed wavefronts. The method proposed here consists
  in an extension of phase diversity to complex amplitude measurements that is
  effective for highly perturbed beams. \requestreviewer{Named} \CAMELOT for Complex
  Amplitude MEasurement by a Likelihood Optimization Tool, it relies on the
  acquisition and processing of few images of the beam section taken along the
  optical path. The complex amplitude of the beam is retrieved from the images
  by the minimization of a Maximum a Posteriori error metric between
  the images and a model of the beam propagation. The analytical formalism of
  the method and its experimental validation are presented. The modulus of the
  beam is compared to a measurement of the beam profile, the phase of the beam
  is compared to a conventional phase diversity estimate. The precision of
  the experimental measurements is investigated by numerical simulations.
\end{abstract}



\renewcommand{\refname}{References and links}

\section{Introduction}

The optical quality of the beam is a critical issue for intense lasers: a good
optical quality is necessary to optimize stimulated amplification in optical
amplifiers, to prevent optical surfaces from deteriorating due to hot spots,
and to optimize the flux density at focus. For these reasons the optical
quality of the beams of intense lasers must be monitored. \requestreviewer{To this end},
wavefront analysis has to cope with the presence of high spatial frequencies
in the beam profile due to \requestreviewer{speckle} patterns. Commonly used
wavefront sensors, whether Shack-Hartmann~\cite{Shack-71b} (SH) or shearing
interferometers~\cite{Primot-a-93}, rely on the assumption of a continuous
phase structure. In order to measure high spatial frequencies, the sampling of
the wavefront has to be fine. This requires additional optical components and
many measurement points, which are manageable assuming large focal plane
sensors. Moreover, these concepts require the reconstruction of the wavefront
from wavefront gradient measurements. \requestreviewer{The reconstruction
  process is only valid if the wavefront is continuous and can be measured on
  a domain that is a connected set in the topological sense.}

To bypass the above limitations, far field wavefront sensing techniques such
as Phase Diversity are an appealing alternative. Phase Diversity, \ie the
recovery of the field from a set of intensity distributions in planes
transverse to propagation, is a well-established method. The vast majority of
the work on this technique has concentrated on the estimation of the phase for
applications related to imaging, assuming that the modulus is known,
as is often reasonable in astronomy at least---see in
particular~\cite{Misell-a-73a,Gonsalves-a-82} for seminal
contributions,~\cite{Mugnier-l-06a} for a review on \PD,
and~\cite{Sauvage-a-07} for an application of \PD to reach the diffraction
limit with Strehl Ratios as high as 98.7\%.
	 
Yet, with intense lasers, the wave profile may present strong spatial
variations~\cite{Fourmaux-a-08}. For this reason, in this application of \PD
one needs to estimate the complex amplitude. Early work on phase and modulus
estimation was performed for the characterization of the Hubble Space
Telescope (HST). Roddier~\cite{Roddier-a-93b} obtained non-binary pupil
modulus images using an empirical procedure combining several Gerchberg-Saxton
(GS) type algorithms~\cite{Gerchberg-74}, while Fienup~\cite{Fienup-a-93b}
used a metric minimization approach~\cite{Fienup-a-93} to estimate the binary
pupil shape, parametrized through the shift of the camera obscuration. 

\requestreviewer{The GS algorithm belongs to a larger class of methods based
  on successive mathematical projections, studied in~\cite{Stark-87}.}
Although there is a connection between projection-based algorithms and the
minimization of a least-square functional of the unknown
wavefront~\cite{Fienup-a-82}, the use of an explicit metric to be optimized is
often preferable to projection-based algorithms for several reasons. Firstly,
it allows the introduction of more unknowns (differential tip-tilts between
images, a possibly extended object, etc); secondly, it allows the
incorporation of prior knowledge about the statistics of the noise and/or of
the sought wavefront; thirdly, projection based algorithms are often prone to
stagnation.

The first work on estimating a wave complex amplitude from phase diversity
data with an explicit metric along with experimental results was published by
Jefferies~\cite{Jefferies-a-02}, in view of incoherent image reconstruction in
a strong perturbation regime. The authors encountered difficulties in the
metric minimization, which may in part be due to the parametrization of the
complex amplitude with its polar form rather than its rectangular one.
\requestreviewer{Experimental results in a strong perturbation regime have also been obtained
by Almoro \etal~\cite{Almoro-a-06} using a wave propagation-based algorithm.
 This algorithm, which can be interpreted as successive mathematical
projections~\cite{Stark-87}, required many measurements recorded at
axially-displaced detector planes. The said technique was adapted for smooth
test object wavefronts using a phase diffuser and, for single plane detection,
with a spatial light modulator\cite{Agour-a-12}.} More recently, Thurman and
Fienup~\cite{Thurman-a-09} studied, through numerical simulations, the
influence of under-sampling in the reconstruction of a complex amplitude from
phase diversity data with an explicit metric optimization, in view of
estimating the pupil modulus and aberrations of a segmented telescope, but
this metric was not derived from the data likelihood.

This paper aims at presenting a likelihood-based complex amplitude retrieval method
 relying on phase diversity and requiring few images, together with an
experimental validation. As in~\cite{Thurman-a-09}, the complex amplitude is
described by its rectangular form rather than the polar one to make the
optimization more efficient. 
The method, named \CAMELOT for \textit{Complex Amplitude Measurement by a
  Likelihood Optimization Tool}, is described in Section~\ref{sec-method}. In
Section~\ref{sec-manip} we validate it experimentally and assess its
performance on a laboratory set-up designed to shape the complex amplitude of
a laser beam and record images of the focal spot at several longitudinal
positions. In particular, a cross-validation with conventional \PD is presented. Finally,
experimental results are confronted to carefully designed simulations taking
into account many error sources in Section~\ref{sec:perf_analysis}. In
particular, the impact of photon, detector and quantization noises on the
estimation precision is studied.

\section{\CAMELOT}
\label{sec-method}

\subsection{Problem statement}
\label{sec-statement}

\begin{figure}[!t]
\centering
\includegraphics[width=0.8\linewidth]{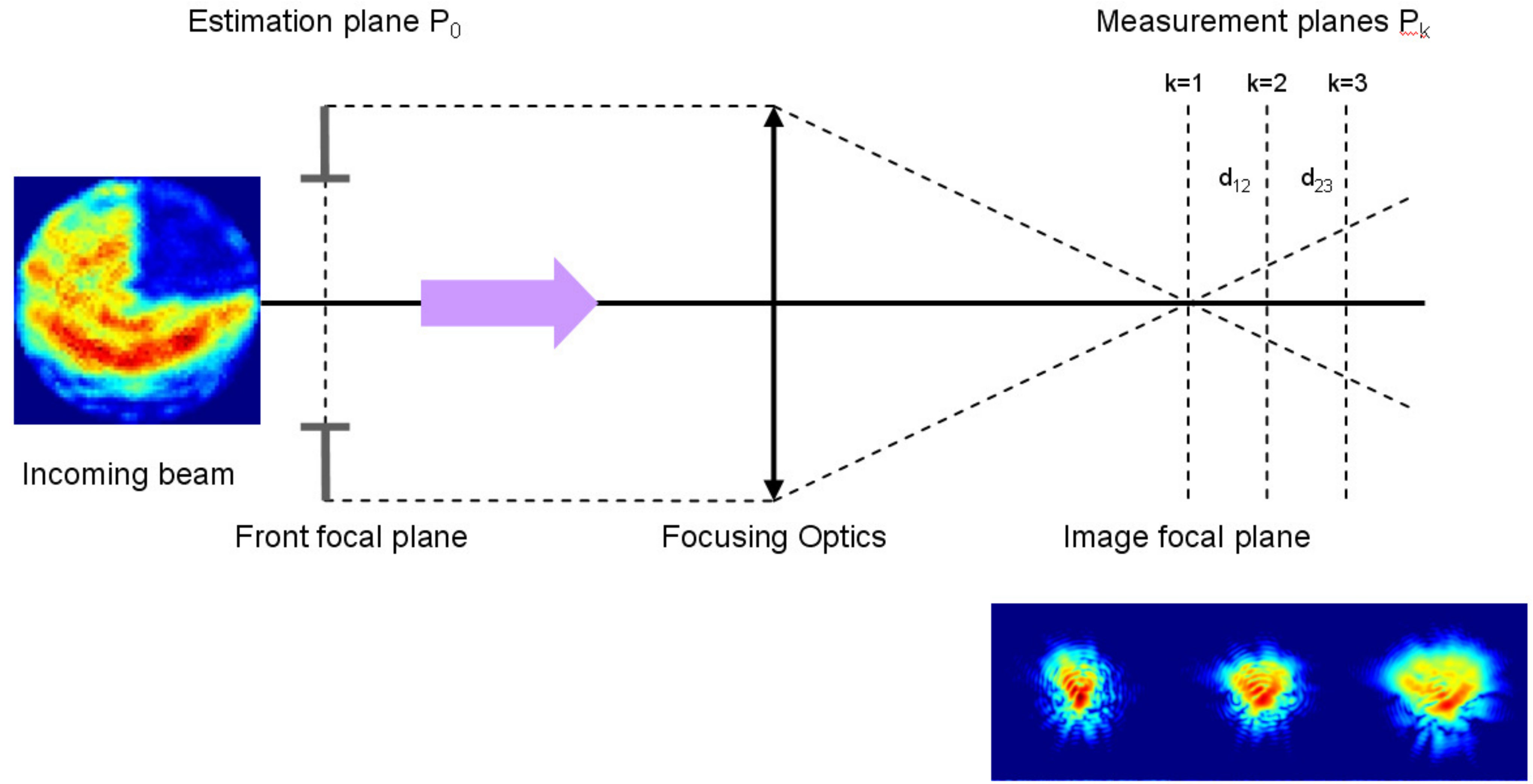}
\caption{Schematic diagram for phase diversity measurement.}
\label{mod_dir}
\end{figure}

The schematic diagram for \PD measurement is presented on Fig.~\ref{mod_dir}.
An imaging sensor is used to record the intensity distributions. As with any
wavefront sensor, the laser beam is focused by optics in order to match the
size of the beam with that of the sensor. With \PD, the sensor area is
installed at the image focal plane of a lens. Note that it is advisable to
install a clear aperture at the \requestreviewer{front} focal plane of the optics in order to
ensure a correct sampling of the intensity distributions. The focal length of
the optics and the diameter of the aperture are chosen in order to satisfy the
Shannon criterion with respect to the pixel spacing.
	
In order to retrieve the complex amplitude of an electromagnetic field, the
relationship between the unknowns and the measurements must be described
mathematically. This description is called the image formation model, or
direct model. 

Let $\Psi_k$ denote the complex amplitude in plane $P_k$. $\Psi_k$ is decomposed onto a finite orthonormal
spatial basis with basis vectors $\{b_{j,k}(x,y)\}_{j=[1,N_k]}$:
\begin{equation}
  \Psi_k(x,y)=\sum_{j=1}^{N_k} \psi_{j,k} b_{j,k}(x,y).
\end{equation}
The coefficients of this decomposition are stacked into  a single column
vector, denoted by $\psi_{k}=[\psi_{j,k}]_{j=[1,N_k]}\in
\complex^{N_k}$. In the following, a pixel basis is used without loss
of generality. 

The field complex amplitude in the plane of the above-mentioned clear
aperture, $P_0$, is supposed to be the unknown. $P_0$ is called hereafter the
estimation plane. We assume that the \PD is performed by measuring intensity
distributions in $N_P$ planes, perpendicular to the propagation axis. $P_k$
($1\leq k\leq N_P$) refer to these planes. The transverse intensity
distributions of the field are measured by translating the image sensor along
the optical axis around the focal plane. The measured signal in plane $P_k$ is
a two dimensional discrete distribution concatenated formally into a single
vector of size $N_k$ denoted by $i_k$. As the detection of the images is
affected by several noise sources, denoting $n_k$ the noise vector, the direct
model reads:
\begin{equation}
\label{eq:i_k}
  i_k={|\psi_{k}|^2} + n_k,
\end{equation}
where $|X|^2 = X\odot X^*$ and $\odot$ represents a
component-wise product. The component-wise product of two complex column
vectors of size $N$ denoted $X=[X_j]^T_{j=[1,N]}$ and $Y=[Y_j]^T_{j=[1,N]}$ is defined as
the term by term product of their coordinates:
\begin{equation}
X\odot Y =[\left(X_{j}Y_{j}\right)]^T_{j=[1,N]},
\end{equation}
In Eq.~(\ref{eq:i_k}), the spatial integration of the intensity
distribution by the image sensor is not taken into account. In practice,
this assumption will remain justified as long as the spatial sampling rate
exceeds the Shannon criterion.

Each $\psi_k$ can be expressed as a linear transformation (a
transfer) of $\psi_0$ and therefore be described by the product of the
propagation matrix $M_k\in\complex^{N_0 \times N_k}$ by $\psi_0$.

Unfortunately, the transverse registration of the different measurements is
experimentally difficult to obtain with accuracy. In order to take these
misalignments into account, differential shifts between planes are introduced
in the direct model via the dot product of $\psi_0$ by a differential shift
phasor $s_k$:

\begin{equation} \psi_k=M_k[\psi_0\odot s_k].
	\end{equation}
The $k$-th differential shift phasor is decomposed on the Zernike tip and tilt
polynomials $Z_2$ and $Z_3$ expressed in the pixel basis~\cite{Noll-a-76},
$a_{2,k}$ and $a_{3,k}$ being their respective coefficients:
\begin{equation} s_k=e^{i(a_{2,k} Z_2+a_{3,k}Z_3)}.
\end{equation}

The misalignment vector $a$ is defined as: $a=\{a_{i,k}\}$. Without loss of
generality, the first measurement plane (k=1) is chosen as the reference plane, so that $a_{2,1}=a_{3,1}=0$.

Finally the image formation model is:
\begin{equation}
  \label{eq:mod_direct}
  i_k = {|M_k[\psi\odot s_k]|^2} + n_k.
\end{equation}

	 The propagation from the estimation plane down to the reference plane is
simulated numerically using a discrete Fourier
transform (DFT), $\Psi_1$ being the far field of $\Psi_0\odot s_1$:
	\begin{equation} \psi_1= \DFT{\psi_0\odot s_1}
	\end{equation} 

The propagation between the reference plane ($k=1$) and plane $P_{k>1}$ is
simulated by a Fresnel propagation performed in Fourier space:
\begin{equation}
  \psi_{k>1}=\frac{\exp{(i\frac{2\pi}{\lambda d_{1k}})}}{i \lambda d_{1k}} \IDFT{\DFT{\psi_1}\odot\exp{(i\pi\lambda d_{1k} \nu^2)}}.
  \end{equation} 
where $d_{1k}$ is the distance between plane $1$ and plane
$k$, $\nu$ is the norm of the spatial frequency vector in discrete Fourier
space, and IDFT the inverse of the DFT operator.

\subsection{Inverse problem approach}
\label{criterion}

In order to retrieve an estimation of $\psi_0$ from the set of measurements,
$\{i_k\}$, the basic idea is to invert the image formation model, \ie the
direct model. For doing so, we adopt the following maximum a Posteriori (\MAP)
framework~\cite{Idier-l-08}: the estimated field $\hat{\psi_0}$ and misalignment coefficients
$\hat{\nv{a}}$ are the ones that maximize the conditional probability of the
field and misalignment coefficients given the measurements, that is the
posterior likelihood $P(\psi_0,\nv{a}|\nv{\{i_k\}})$.
According to Bayes' rule:
	\begin{equation}
      \label{eq:vraisemblance}
	P(\psi_0,\nv{a}|\{i_k\})\propto P(\{i_k\}|\psi_0,\nv{a})P(\psi_0)P(\nv{a})
	\end{equation}
	 where $P(\psi_0)$, respectively $P(a)$, embody our prior knowledge on $\psi_0$,
     respectively $a$. 

	The MAP estimate of the complex
    field corresponds to the minimum of the negative logarithm of the
    posterior likelihood:
	\begin{equation}
	\label{eq:argmin}
	(\hat{\psi_0},\hat{\nv{a}}) = \arg\min_{\psi,\nv{a}} J(\psi_0,\nv{a}) 
	\end{equation}
	which, under the assumption of Gaussian noise, takes the following form:
	\begin{equation}
	\label{eq:J}
	J(\psi_0,\nv{a}) = \sum_{k=1}^{N} \frac{1}{2}( \nv{i_k} -\modsqhada{M_k[\psi_0 \hadamard
      \nv{s_k}]})^T C_k^{-1} (\nv{i_k}
    -\modsqhada{M_k[\psi_0\hadamard\nv{s_k}]}) -\ln P(\psi_0) - \ln P(a) ,
	\end{equation}
where	$C_k = \langle \nv{n_k} \cdot \nv{n_k}^T \rangle$ is the covariance
matrix of the noise on the pixels recorded in plane $k$ (diagonal if the noise
is white) and $-\ln P(\psi)$ and $-\ln P(a)$ are
regularization terms that embody our prior knowledge on their arguments.

In the experiment and in the simulations presented hereafter, we have not
witnessed the need for a regularization, so in the following we shall take
$P(\psi)=P(a)=$constant, and the MAP metric of Eq.~(\ref{eq:J}) reduces to a
Maximum-Likelihood (ML) metric.

This metric is similar to but different from the intensity criterion suggested by Fienup
in~\cite{Fienup-a-93}. Indeed, in our likelihood-based approach, $C_k^{-1}$
enables us to take into account not only bad pixels (either saturated or dead)
but also noise statistics.

\subsection{Minimization}
\label{minimization}
	
In Eq.~(\ref{eq:argmin})~$J$ is a non-linear \requestreviewer{real-valued} function of $\Psi_0$ and $\nv{a}$.
In order to perform its minimization a method based on a quasi-Newton
algorithm, called variable metric with limited memory and bounds
method (VMLM-B)~\cite{Thiebaut-p-02} is used. The VMLM-B method requires the
analytical expression of the gradient of the criterion. \requestreviewer{The complex gradient
of $J$ with respect to $\psi$~\cite{matrixcookbook,Kreutz-Delgado-a-09},
denoted $\nabla J(\psi)$, is defined in Eq.~(\ref{nabla_JE_1}) as the complex
vector having the partial derivative of $J$ with respect to $\Re (\psi)$
(respectively $\Im (\psi)$) as its real (respectively imaginary) part.} 

\begin{equation}
\label{nabla_JE_1}
\nabla J(\psi)=\derpar{J}{\Re (\psi)} + j \derpar{J}{\Im (\psi)} =
-\sum_{k=1}^{N}2 \underbrace{\nv{s_k}^*\hadamard M_k^{*^T}}_{4}\left( \underbrace{C_k^{-1}}_{3}\left[(\underbrace{M_k [\psi\hadamard \nv{s_k}]}_{2})\hadamard ( \underbrace{\nv{i_k}-|M_k [\psi\hadamard\nv{s_k}]|^2)}_{1}\right]\right).
\end{equation}

	Four factors can be identified in Eq.~(\ref{nabla_JE_1}), which allows a
    physical interpretation of this somewhat complex expression: 
	\begin{enumerate}
	\item the computation of the difference between measurements and the direct model;
	\item weighing this difference by the result of the direct model;
	\item whitening the noise to take into account noise statistics;
	\item a reverse propagation  that enables the projection of the gradient
      into the space of the unknowns and takes into account differential tip/tilts. 
	\end{enumerate} 
	
	The minimization of $J$ must also be performed with respect to misalignment coefficients. The following analytical expression has been obtained and implemented:
	\begin{equation}
	\label{nabla_grad_ak} 
	\derpar{J}{a_{i,k}}=-\Im\Brack{\Paren{\psi\hadamard\nv{s}_k\hadamard\nv{Z_i}}^T\Paren{2 M_k^{*^T} C_k^{-1}\left[(M_k [\psi\hadamard \nv{s_k}])\hadamard ( \nv{i_k}-|M_k [\psi\hadamard\nv{s_k}]|^2)\right]}^*}.
	\end{equation}

\section{Experiment}
\label{sec-manip}

\subsection{Principle of the experiment}
\label{sec-principle}
\begin{figure}[!t]
\centering
\includegraphics[width=\linewidth]{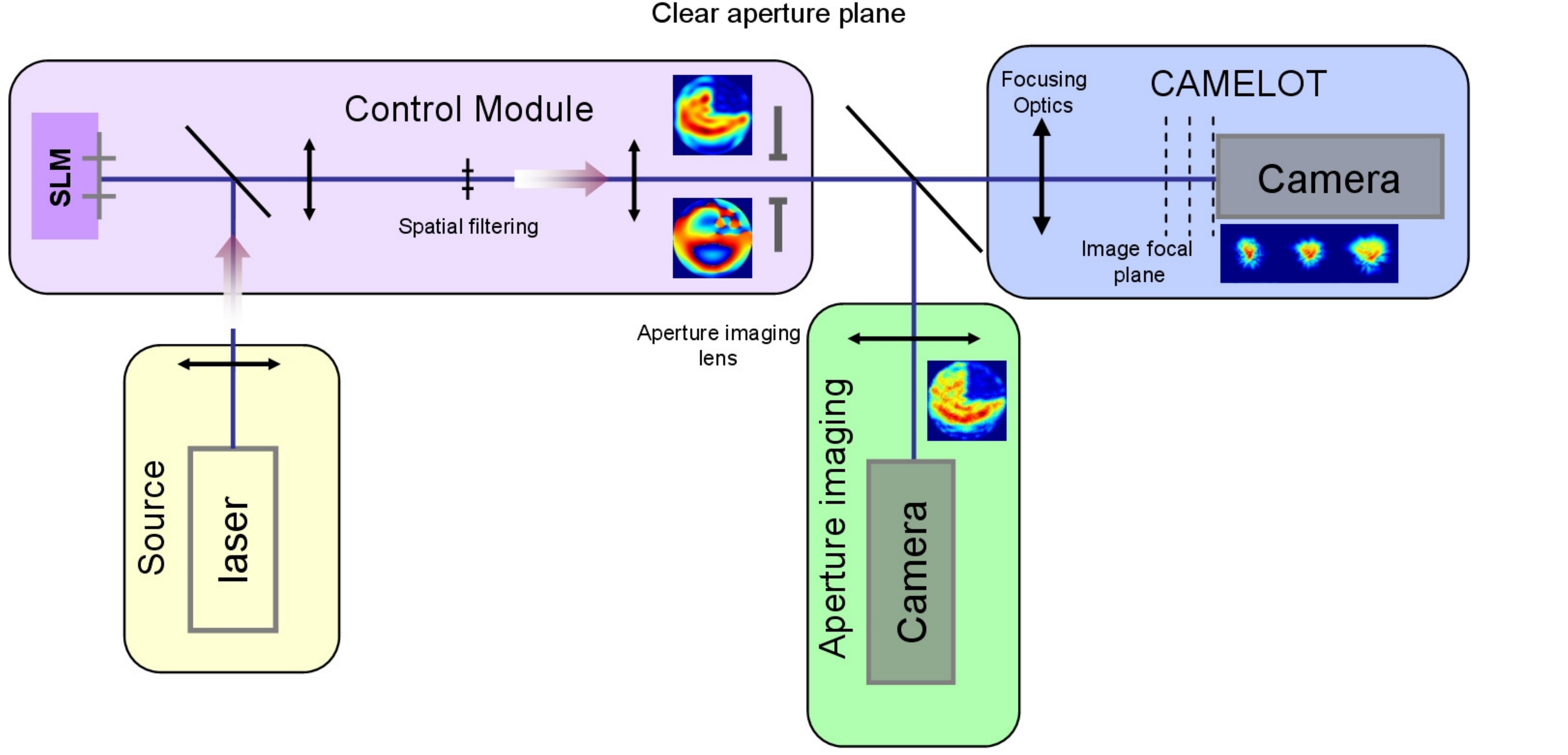}
\caption{Experimental setup for \CAMELOT validation.}
\label{fig_expe_setup}
\end{figure}

The objective of the experiment is to validate \CAMELOT experimentally on a
perturbed laser beam. \requestreviewer{The experiment is performed using a low power continuous
fibered laser whose complex amplitude is modulated to create spatial
perturbations.} The experimental setup is illustrated on
Fig.~\ref{fig_expe_setup}.
The spatial modulation of the laser beam (phase and
modulus) is controlled by a field control module and conjugated with the clear
aperture plane.

The beam going through the clear aperture crosses a beam splitter and is
focused on the \CAMELOT camera of Fig.~\ref{mod_dir}. The camera is used to
record three near focal plane intensity distributions, which are given as
inputs to \CAMELOT. The field estimated by the latter will be denoted
$\psi_c=A_Ce^{i\varphi_C}$ in the following. The beam reflected on the
beam-splitter is used to record the intensity distribution $I_M$ with an image
sensor conjugated with the aperture plane. 

\CAMELOT's estimation of the complex field is cross-validated in the two
following ways: regarding field modulus, $A_C$ is compared with $A_M$, called
measured modulus of the field hereafter, which is computed as the square-root
of image $I_M$: $A_M=\sqrt{I_M}$.

Regarding phase, \CAMELOT's estimate is compared to the result of a
straightforward adaptation of a classical \PD algorithm~\cite{Mugnier-l-06a}
that will be referred to as conventional \PD. Phase diversity is now a well
established technique: it has been used successfully for a number of
applications~\cite{Mugnier-l-06a}, including very demanding ones such as
extreme adaptive optics~\cite{Sauvage-a-07,Paul-a-13}, and its performance has
been well characterized as a function of numerous factors, such as system
miscalibration, image noise and artefacts, and algorithmic
limitations~\cite{Blanc-a-03}.
In classical \PD, only the pupil phase is sought for, while the pupil
transmittance is supposed to be known perfectly\requestreviewer{. This is
  probably the main limitation of this method, regardless of how the
  transmittance is known, whether by design or by measurement. In the case of
  a transmittance measurement, a dedicated imaging subsystem must be used,
  hence increasing the complexity of the setup. Misalignment issues (scaling,
  rotation of the pupil) have to be addressed due to the potential mismatch
  between the true pupil (which gives rise to the recorded image) and the pupil that is
  assumed in the image formation model.}
In this paper, for conventional \PD, the pupil transmittance is set to the
measured modulus $A_M$ and the data used are two of the three near focal plane
intensity distributions: the focal plane image and the first defocused image.


\subsection{Control of the field}
\label{sec:control}
\requestreviewer{In this section, we present the modulation method used to control the field
and how the result of the modulation is measured in the aperture plane.} 

The modulation of the complex amplitude is obtained using the field control
method suggested by Bagnoud~\cite{Bagnoud-a-04}: a phase modulator is followed
by a focal plane filtering element.

The phase modulation is performed with a phase only SLM (Hamamatsu LCOS SLM
x10-468-01) with 800x600 $20\mum$ pixels. \requestreviewer{The laser source is a $20\mW$
continuous laser diode at $\lambda = 650\nm$ injected into a $4.6\mum$ core
single-mode fiber. At the exit of the fiber, the beam is collimated, linearly
polarized and then reflected on the SLM.} The clear aperture diameter is
$D=3\mm$ on the surface of the SLM. It is conjugated with a unit magnification
onto the clear aperture plane. The spatial modulation and filtering are
designed to control 15$\times$15 resolution elements in the clear aperture \ie
15 cycles per aperture. The effective result of the control in the clear
aperture plane is called the true field. Its modulus is denoted $A_T$.
	
\begin{figure}[htbp!]
 \centering
\begin{tabular}{*{1}{@{\,}c}}
 \includegraphics[width=0.8\linewidth]{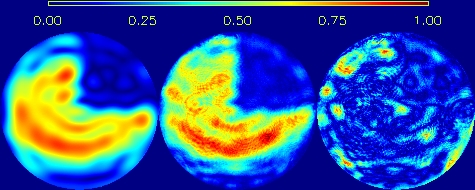}
\end{tabular}
\caption{Left: modulus of the simulated field ($A_S$), center:
   measured modulus ($A_M$), right:  $2.5\times|A_M -A_S|$.}
 \label{fig_control}
\end{figure}

\begin{figure}[htbp!]
 \centering
 \includegraphics[width=0.25\linewidth]{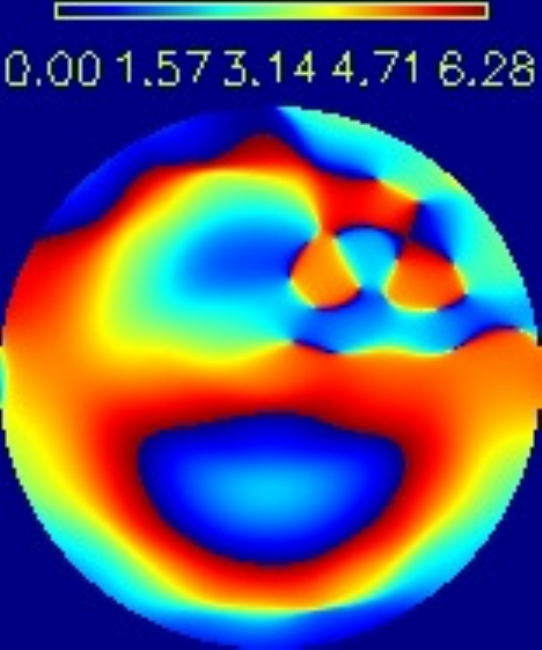}
\caption{Phase of the simulated field.}
 \label{fig_phase}
\end{figure}

The spatial modulation and filtering have been modeled by means of an
end-to-end simulation. The result of the simulation is
denoted by $\psi_S$. Its modulus, denoted $A_S$, is presented on the left of
Fig.~\ref{fig_control}. It shows smooth variations as can be typically
observed on an intense pulsed laser~\cite{Fourmaux-a-08}. It has been
truncated in the upper right corner to simulate a strong vignetting effect.
The phase of the field is presented on Fig.~\ref{fig_phase}. It is dominated
by a Zernike vertical coma ($Z_7$) of $11~ \rad$ \requestreviewer{Peak-to-Valley} (PV), typical of a strong
misalignment of a parabola for instance.

The modulus of the true field is measured with the aperture plane imaging
camera (see Fig.\ref{fig_expe_setup}). The image is recorded with a high
spatial resolution (528 pixels in a diameter) and with a high SNR per pixel
(60 in average). The result of the measurement,$A_M$, is
presented at the center of Fig.~\ref{fig_control}.

In order to quantify the proximity of the simulated modulus $A_S$ with the
measured one $A_M$, the following distance metric is defined:
 \begin{equation}
 \label{eq:metrica}
 \ert_{MS}=\frac{ \sum_{j=1}^{N_0}\modcsq{A_{M,j}- A_{S,j}}}{\sum_{j=1}^{N_0} \modcsq{A_{S,j}}}.
 \end{equation}

 To enable their comparison, $A_S$ and $A_M$ have been normalized in flux
 ($\sum_{j=1}^{N_0} \modcsq{A_{S,j}}=\sum_{j=1}^{N_0} \modcsq{A_{M,j}}=1$). 

 The spatial distribution of the modulus of the difference between $A_S$ and
 $A_M$ is presented on the right of Fig.~\ref{fig_control}. It has been
 multiplied by a factor 2.5 to present a dynamic comparable to $A_M$. 

 The corresponding distance $\ert_{MS}$ is found to be $\ert_{MS}=0.044$, whereas
 in the case of a photon noise limited measurement simulations show that one
 should rather obtain $\ert_{MS}=5~10^{-5}$.

 A preliminary Fourier-based analysis of the difference between $A_S$ and
 $A_M$ enables to identify a clear cut separation between spatial frequencies
 below the cut-off frequency of the spatial modulation (15 cycles per
 aperture) and higher spatial frequencies. Due to the spatial filtering by the
 control module, the low spatial frequencies part of the difference between
 $A_S$ and $A_M$ can be attributed mostly to model errors between the simple
 simulation performed to compute $A_S$ and the effective experimental setup of
 the control module. For instance in the simulation, the spatial filter is
 assumed to be centered on the optical axis, the SLM illumination is assumed
 to be perfectly homogeneous, and lenses and optical conjugations are supposed
 to be perfect. This part of $\ert_{MS}$ is evaluated to $0.04$, that is more
 than $80\%$ of the total. The high frequencies part of $\ert_{MS}$ comes from
 optical defects on relay optics and noise influence on $A_M$ measurement. It
 is evaluated to $4~10^{-3}$. This gives insight on the measurement
 precision of $A_M$, which should thus be of the order of $4~10^{-3}$.

\subsection{\CAMELOT estimation}
\label{practical_implementation}
\subsubsection{Practical implementation of phase diversity measurement}
The intensity distributions in the different planes are recorded by
translating the sensor along the optical axis around the focal plane.
\requestreviewer{The amount of defocus must be large enough to provide
  significantly different intensity distributions so as to facilitate the estimation
  process. However, small translation distances are preferred for the sake of
  experimental easiness. For conventional \PD a defocus of $\lambda$ PV is
  often chosen as it maximizes the difference between the focal plane
  intensity distribution and the defocused image.} Therefore, the amplitude of
defocus between the position of each plane is fixed to $\lambda$ PV.
Considering such a defocus, it appears that no less than three different
measurement planes are required with \CAMELOT. The first image is located in
the focal plane, the second one is at a distance corresponding to $\lambda$ of
defocus PV, the third one at a distance of 2$\lambda$ PV from the focal plane.
\requestreviewer{The relation between the PV optical path
difference $\delta_{OPD}$ in the aperture plane and the corresponding translation distance between two successive
planes $d_{kk+1}$ is:
\begin{equation}
d_{kk+1}= 8 (f/D)^2\delta_{OPD}.
\end{equation}
For the experiment, the focusing optics focal length is $f=100\mm$ and the
aperture diameter is $D=3\mm$. Consequently the translation distances are
$d_{12}=d_{23}=5.78\mm$ between the successive recording planes.}

The intensity distributions are recorded with a Hamamatsu CMOS Camera (ORCA
$R^2$). Its characteristics are the following: a pixel size and spacing $
s_{pixel}=6.452\mum$, a readout noise standard deviation
$\sigma_{ron}=6~e^-$\,rms, a full well capacity of $18 000~e^-$, and a $12$
bit digitizer.

As explained in Section \ref{sec-statement}, the sampling must fulfill the
Shannon criterion. The focusing optics focal length and the aperture diameter
combined with the pixel size lead to a theoretical Shannon oversampling factor
$\lambda f/(2s_{pixel}D) = 1.68$ for $\lambda = 650\nm$.

The total number of photo-electrons must be large enough to prevent the
estimation from being noise limited. Simulations of the system that will be
detailed in Section~\ref{sec:perf_analysis} demonstrate that
$N_{phe}=5\times10^7$ total photo-electrons are sufficient. Due to the limited
full well capacity of the sensor, $p = 10$ short exposures images are added to
reach this number. For each image, background influence has been
removed by a subtraction of an offset computed from the average of pixels
located on the side of the images (hence not illuminated).

The noise covariance matrix $C_k$ is approximated by a diagonal matrix whose
diagonal terms, $[C_k]_{jj}$, correspond to the sum of the variances of the
photon noise, read out noise and quantization noise:
 \begin{equation}
 [C_k]_{jj} = [\sigma^2_{ph,k}]_j + p(\sigma^2_{ron} + q^2/12)
 \end{equation}
where $q$ is the quantization step. In practice, the readout noise variance
map $\sigma^2_{ron}$ is calibrated beforehand and the photon noise variance
$\sigma^2_{ph,k}$ is approximated from the image by:
$[\hat{\sigma}^2_{ph,k}]_j=\max([\nv{i}_k]_j,0)$ on pixel $j$~\cite{Mugnier-a-04}.

The three recorded near-focal intensity distributions are presented at the top
of Fig.~\ref{psfs}. The full size of one image is $N_{pix}=214\times214$ pixels.
For the figure, a region of interest of $140\times140$ pixels, centered on the
optical axis, is selected. From left to right the focal plane image, the first
defocused image and the second one are displayed. The colorscale is
logarithmic.

\subsubsection{Results}

The three focal plane images are now used in the \CAMELOT estimation. The
minimization process is initiated with a homogeneous modulus field, a phase
set to zero and differential tip/tilts also set to zero. The number of
estimated points in the estimation plane is $N_0=64\times 64$. The
current implementation of the algorithm is written in the IDL language.

\begin{figure}[!t]
\centering
\includegraphics[width=0.8\linewidth]{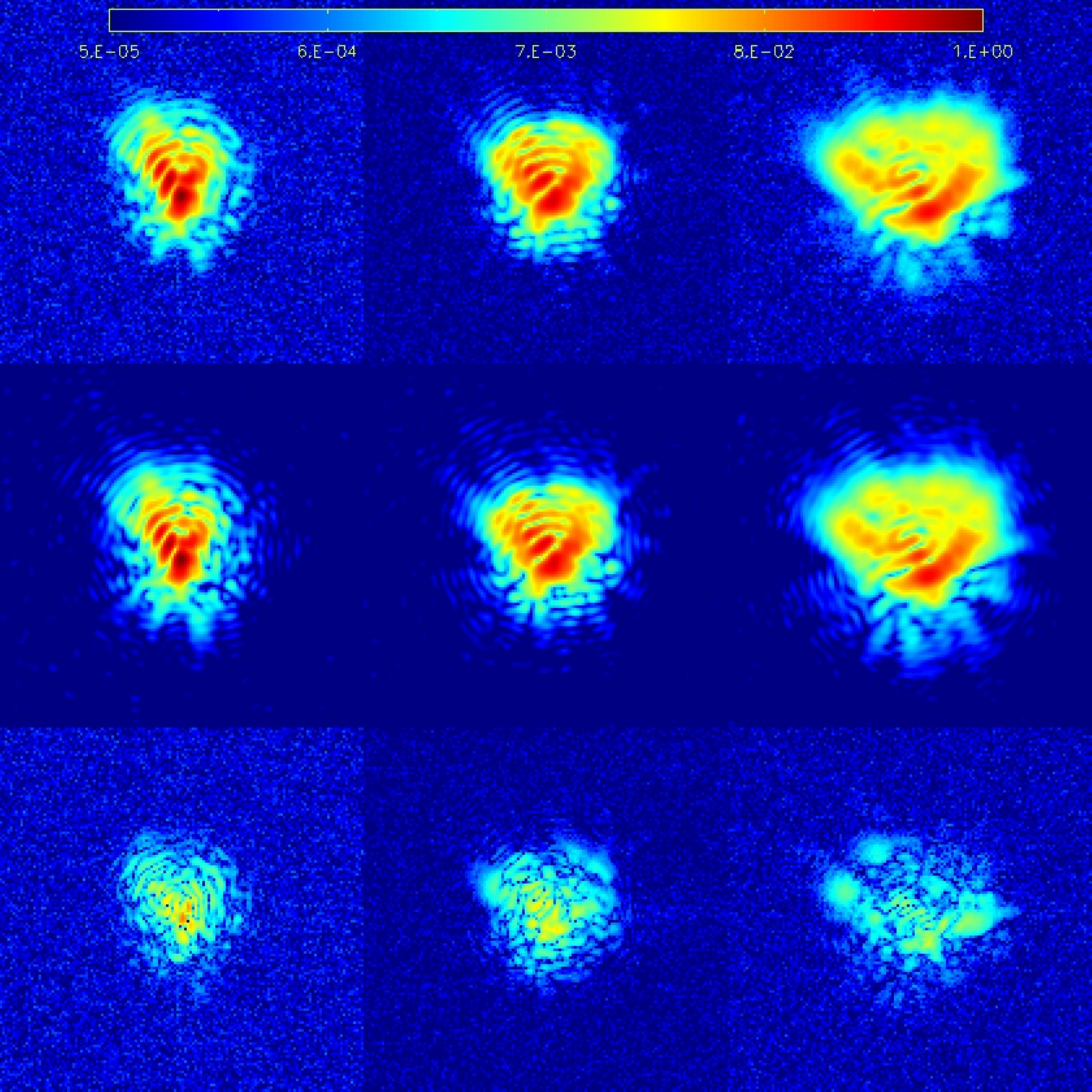}
\caption{Left: focal plane images ($k=1$); center: first defocused images
  ($k=2$); right: second defocused image ($k=3$). Top: experimental images;
  middle: direct model result for $\modsqhada{M_k\psi_c}$; bottom: residuals, \ie
  $\modc{\nv{i_k} -\modsqhada{M_k\psi_c}}$. Logarithmic scale, $140\times140$ pixels ROI centered on the optical axis.}
\label{psfs}
\end{figure}

A relevant measurement of the success of our inversion method is the quality
of the match between the data $i_k$ and the model that can be computed from
the estimated field through Eq~(\ref{eq:mod_direct}). This latter map is presented on the
middle row of Fig.~\ref{psfs}. The moduli of the differences between the
measurements and the direct model, that is to say estimation residuals, are
displayed on the bottom row of the same figure. These residuals are below
$1\%$ of the maximum of the measurements on the three planes.

The estimated shifts  are $(-0.73, -0.56)$ and $(0.97, 0.61)$ pixel for x and y
directions for $P_2$ and  $P_3$ respectively.

\begin{figure}[!t]
\centering
\includegraphics[width=0.8\linewidth]{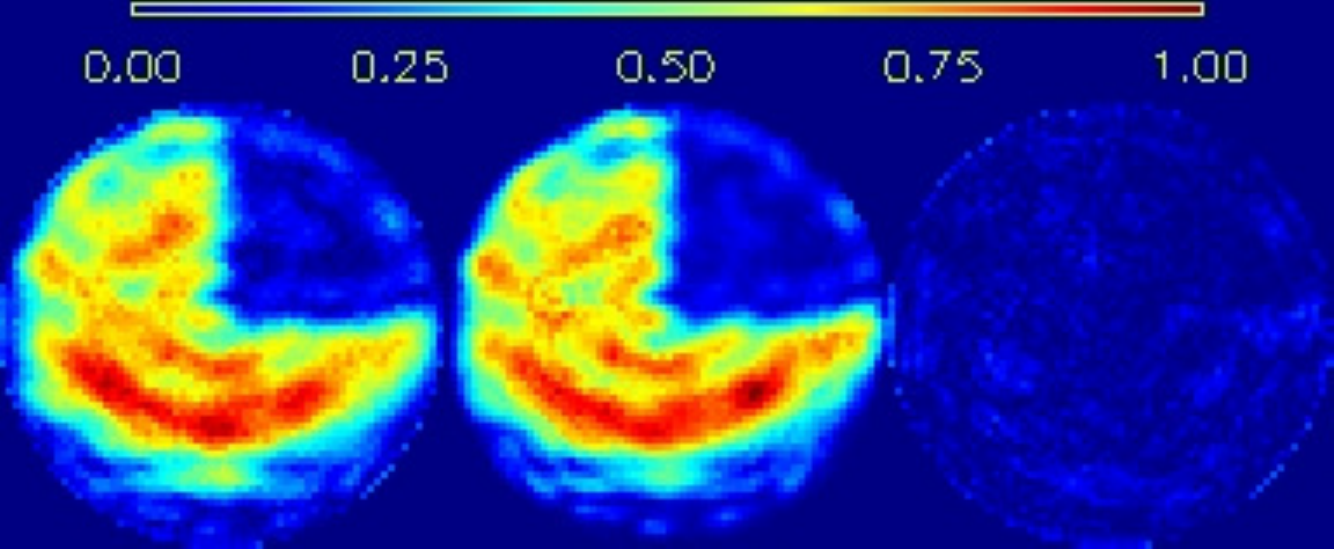}
\caption{From left to right: $A_c$, $A_M$ and $\modc{A_C-A_M}$.}
\label{fig_amp}
\end{figure}

The modulus of \CAMELOT's estimated field, $A_C$, is presented on the left of
Fig.~\ref{fig_amp}. It has been normalized in flux to enable the comparison
with the measured modulus $A_M$, shown in the center of the figure. For the
comparison, $A_M$ has been sub-sampled and resized to $64\time64$ pixels. The
modulus of the difference between $A_C$ and $A_M$ is represented on the right.
The main spatial structures of $A_M$ are well estimated by the method. The
estimation residuals are below $20\%$
of the maximum of the measured modulus, even in the zones where the flux is
low (top right corner). The distance between the two moduli is $\ert_A=0.01$ only.
This must be compared to the error between $A_M$ and $A_S$ reported in
Section~\ref{sec:control} which was found to be $0.044$. \CAMELOT thus 
delivers a modulus estimation that is several times closer to $A_M$ than
$A_S$ is.

\begin{figure}[!t]
\centering
\includegraphics[width=0.8\linewidth]{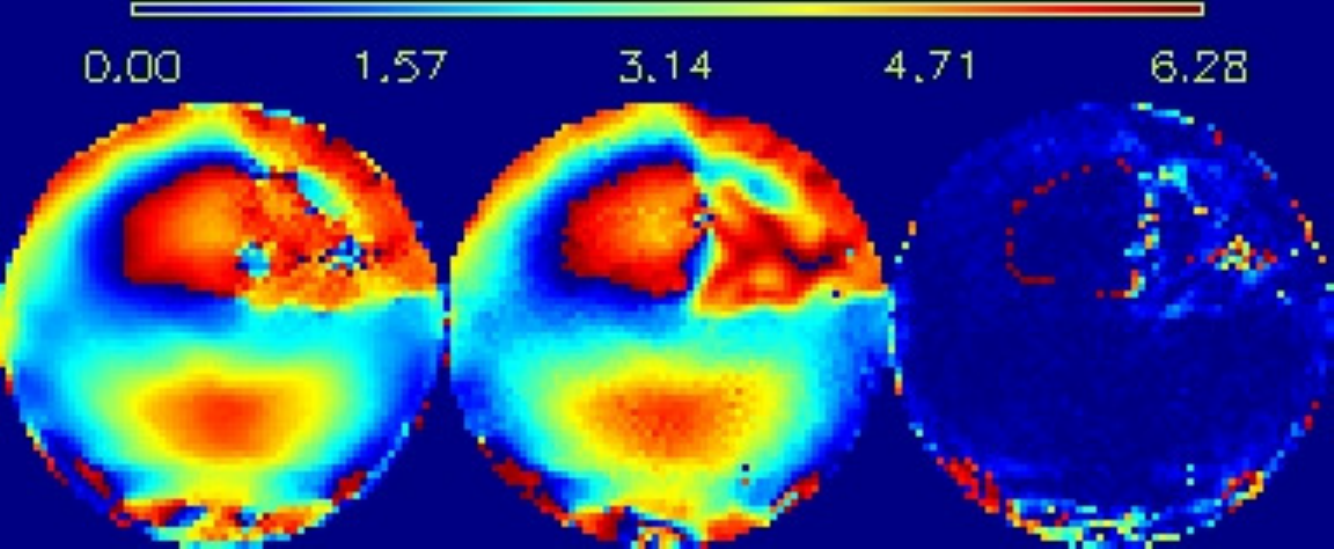}
\caption{From left to right: $\varphi_c$, $\varphi_{PD}$ and $\modc{\varphi_C-\varphi_{PD}}$.}
\label{fig_phase_c}
\end{figure}

We now compare the phase estimated by \CAMELOT to the result of the conventional
\PD method \requestreviewer{described in Section~\ref{sec-principle}}. The comparison is presented on Fig.~\ref{fig_phase_c}. The
phase of $\psi_C$, $\varphi_C$, is presented on the left of the figure, the
phase of the conventional phase diversity,$\varphi_{PD}$, is in the center of
the figure, the modulus of their difference is presented on the right. For
comparison of these phase maps, their differential piston has been set to 0.
In the zones where the modulus is greater than $10\%$ of the maximum modulus
of $\psi_M$, the maximum of the phase residuals is below $2\pi/10$ \rad.

Additionally concerning  conventional phase diversity, setting $A_C$ as the pupil
transmittance instead of $A_M$ enables a better fit to the data with a 5\%
smaller criterion at convergence of the minimization, which is yet another
indicator of the quality of the modulus $A_C$ estimated by \CAMELOT. 

\requestreviewer{\CAMELOT and conventional phase diversity algorithms have
  about the same complexity. The former converges after less than 300
  iterations and requires less than $10\min$ of computation, while the latter
  takes slightly more iterations and time. The main difference between the two
  methods lies in the modeling of the defocus between the measurement planes:
  in \CAMELOT it is performed by a Fresnel propagation, hence requiring two
  additional DFTs for each defocused plane. However, \CAMELOT appears here as
  fast as conventional phase diversity to achieve comparable results. The
  convergence time demonstrated here makes \CAMELOT suitable for the
  measurement and control of slowly varying aberrations such as those induced
  by thermal expansion of mechanical mounts for instance. The use of an
  appropriate regularization metric should contribute to speed up the
  computation. Recent work on real time conventional phase diversity
  demonstrated that few tens of Hertz are achievable with a dedicated (but
  commercial off-the-shelf) computing architecture~\cite{Dolne-a-09}.
  According to the author of the latter reference, several hundreds of Hertz
  could even be achieved in a very near future, making phase diversity
  compatible with the requirements of the measurement and control of
  atmospheric turbulence effects. We believe that these conclusions can be
  generalized to \CAMELOT considering the similar convergence demonstrated
  here compared to conventional phase diversity and the the possibilities
  brought by Graphical Processing Units to speed up Fresnel propagation
  computations~\cite{Nishitsuji-p-11}.}

\section{Performance analysis by simulations}
\label{sec:perf_analysis}

We now analyze the way these results compare with simulations. We simulate
focal plane images from $\psi_C$ while taking into account the main
disturbances that affect image formation: misalignments, photon and detector
noises, limited full well capacity, quantization and miscalibration.
	
For the numerical simulation of the out-of-focus images, a random tip/tilt
phase is added to the field before propagation to the imaging plane in
order to simulate the effect of misalignment. The standard deviation of the
latter corresponds to half a pixel in the focal plane.

Then two cases are considered. The first case, hereafter called perfect
detector, assumes a detector with noise but an infinite well capacity and no
quantization. Each image is first normalized by its mean number of
photo-electrons $N_{phe,k}$, then for each pixel a Poisson occurrence is
computed, and a Gaussian white noise occurrence with variance $\sigma_{ron}^2$
is then added to take into account the detector readout noise. The second case
corresponds to a more realistic detector:

for a given value of the desired  number of photo-electrons $N_{phe,k}$,
the corresponding image is computed as the addition of as many ``short-exposures'' as needed
in order to take into account the finite well capacity, and each of these
short exposures is corrupted with photon noise, read-out noise and a 12~bit
quantization noise. The same number of photo-electrons is attributed to each
image : $N_{phe,k}=N_{phe}/N_P$ where $N_{phe}$ designates the total number of
photo-electrons.

\begin{figure}[!t]
\centering
\includegraphics[width=0.7\linewidth]{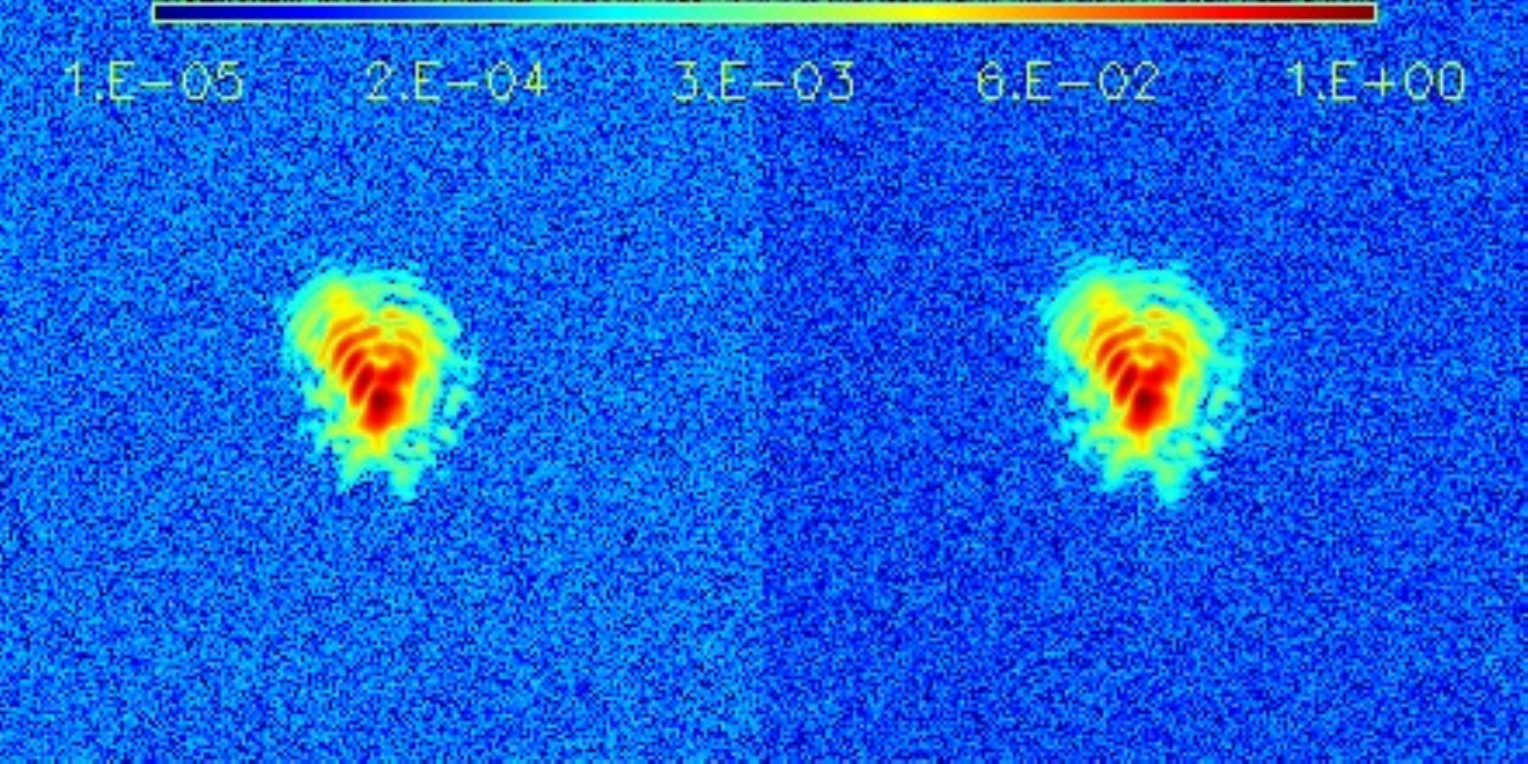}
\caption{Focal plane images, left: simulation ($N_{phe,1}=1.6~10^7$,
  $10$ short exposures), right: experiment.}
\label{fig_psf_dyn}
\end{figure}

The simulated long exposure focal plane image is presented on the left of
Fig.~\ref{fig_psf_dyn}. This image is obtained for $N_{phe,1}=1.6~10^7$
photo-electrons by adding $10$ short exposures. It can
be visually compared with the experimental focal plane image recorded in comparable conditions (right): the similarity
between the two images illustrates the relevance of the image simulation.

\begin{figure}[!t]
\centering
\includegraphics[width=0.8\linewidth]{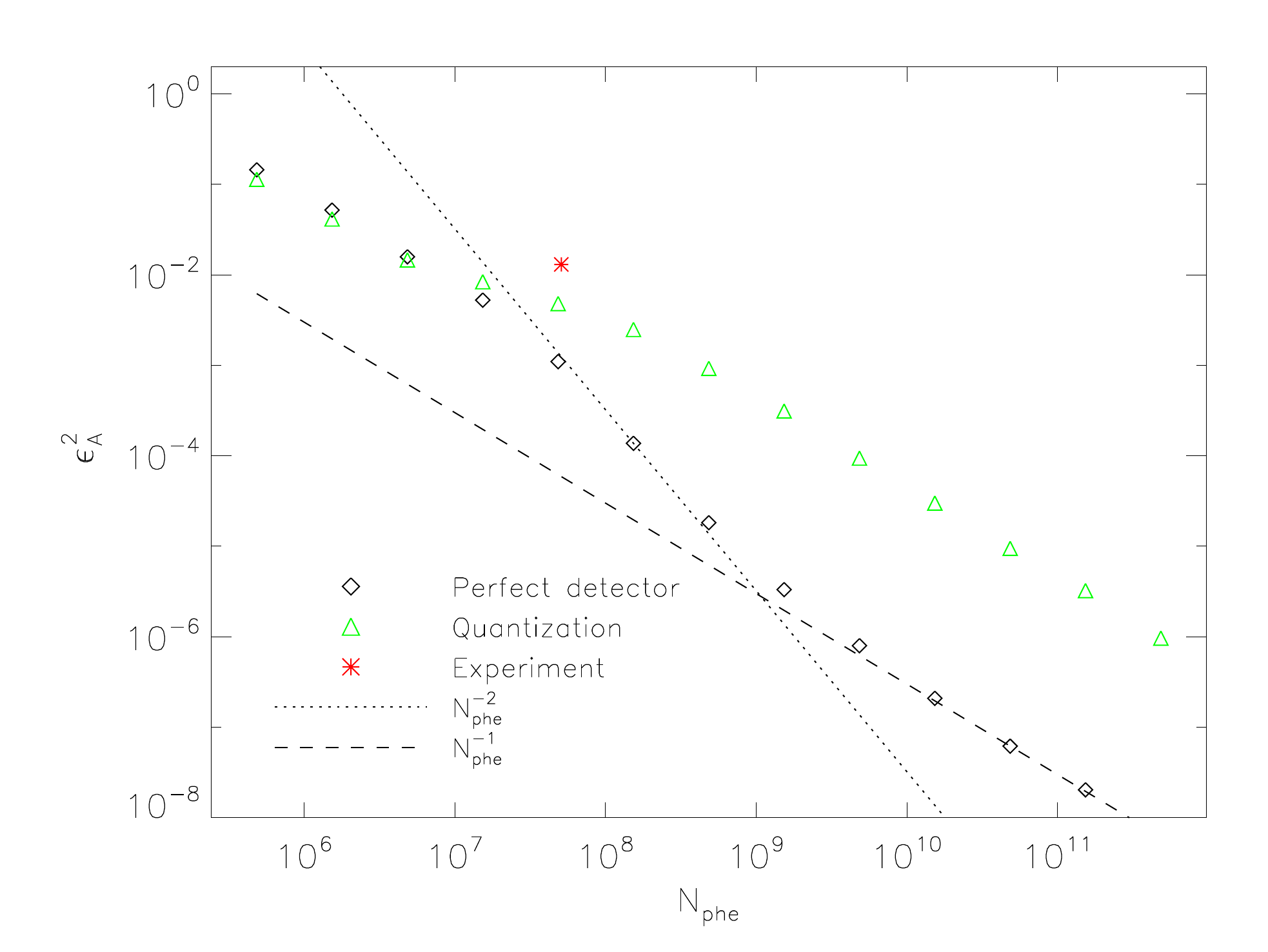}
\caption{$\ert_A$ as a function of average photo-electrons per pixel
  $N_{phe}$. Black diamonds ($\diamond$): perfect detector, green triangles
  ($\triangle$): finite well capacity and quantization, red asterisk ($\ast$):
  experiment result, dotted line: $N_{phe}^{-2}$, dashed line: $N_{phe}^{-1}$.}
\label{fig_varepsilon_a}
\end{figure}

The modulus estimation error $\ert_{A}$ is plotted on Fig.~\ref{fig_varepsilon_a} as a
function of $N_{phe}$ for the two detector cases. Ten uncorrelated occurrences are
averaged to compute $\ert_A$. The result of the comparison of $A_C$ with $A_M$
obtained from the experiment is also reported on the figure (abscissa
$N_{phe}=5~10^7$ photo-electrons). 

For the perfect detector case and $N_{phe}$ small compared to $5~10^8$, $\ert_A$
follows a $N_{phe}^{-2}$ power law. For greater values, the power law turns to
$N_{phe}^{-1}$. This can be explained by the relative weights of the readout
and the photon noise. For $N_{phe}\leq 5~10^8$, the average flux on
illuminated pixels, that can be approximated by $N_{phe}/(N_P~n_i)$ where $n_i$
is the average number of illuminated pixels per image plane ($n_i\approx
100$), is smaller than the total readout noise contribution for one image
plane, that is $N_{pix}^2\sigma_{ron}^2$. Thus, readout noise dominates. For
$N_{phe} \geq 5~10^8$, the photon noise contribution becomes predominant and
the noise propagation of $\ert_A$ follows a $N_{phe}^{-1}$ power law. It is
confirmed here, as stated in Section~\ref{practical_implementation}, that
$\ert_A$ for the experiment is not limited by photon  noise.

The analysis of $\ert_A$ for the case with finite well capacity and quantization
shows that it follows the perfect detector case up to about $N_{phe}=5~10^6$
photo-electrons, then starts to follow a $N_{phe}^{-1}$ power law. $N_{phe}=5~10^6$
corresponds to the flux necessary to saturate the well capacity of the sensor.
Above this limit, images are added to emulate the summation of
``short-exposures'' images. Hence, the noise level in the measurements starts
to depend on the number of summed images, that is to say on $N_{phe}$, with a $N_{phe}^{-1}$ power law.

The error level obtained from the comparison between the modulus measurement
and \CAMELOT estimation ($\ert_{MC}=0.013$) is comparable to the error level
between \CAMELOT estimation and the true modulus in the case of the more
realistic detector ($\ert_A=5~10^{-3}$): they differ approximately by a factor
two. A significant part of this difference comes from the fact that the
measured modulus $A_M$ too is imperfect \ie is only an estimate of the true
modulus $A_T$, due to experimental artefacts, notably differential optical
defects that affect image formation on the aperture plane imaging setup and
noise influence. This claim is also supported
by the fact that the phase estimated by conventional \PD fits the measurements
better when the pupil transmittance is set to the modulus estimated by
\CAMELOT, $A_C$, instead of the measured modulus $A_M$. The Fourier-based analysis mentioned at the end of
Section~\ref{sec:control} delivers an estimate of the measurement precision of
$A_M$ that is evaluated to $4~10^{-3}$. 

As a final remark, one can note that Fig.~\ref{fig_varepsilon_a} can also be
useful to evaluate not only the estimation error on the field modulus but the
total error on the complex field itself.
Indeed, we have calculated that this total error is on average simply twice
the error on the modulus, or $2\ert_A$. This can be particularly useful for
designing a complex field measurement system based on \CAMELOT.

\section{Conclusion}
In this paper, we have demonstrated the applicability of \PD to the
measurement of the phase and the amplitude of the field in view of laser beam
control.
The resolution of the inverse problem at hand has been tackled through a
MAP/ML approach.
An experimental setup has been designed and implemented to test the ability of
the method to measure strongly perturbed fields representative of
misaligned power lasers.
The estimated field has been confronted to measured modulus using pupil
plane imaging and to phases estimated with classical two-plane \PD (using the
measured aperture plane modulus).
It has been shown that the estimation accuracy is consistent with carefully
designed numerical simulations of the experiment, which take into account
several error sources such as noises and the influence of quantization.
Noise propagation on the field estimation has been studied to underline the
capabilities and limitations of the method in terms of photometry. 

Several improvements \requestreviewer{to the method} are currently considered. They include the estimation of
a flux and of an offset per image, adding a regularization metric in the
criterion to be optimized for the reconstruction of the complex field on a
finer grid and speeding up the computations.

\requestreviewer{The work presented in this paper focuses on the case of a
  monochromatic beam. The application of \CAMELOT to the control of intense
  lasers with a wider spectrum, as it is the case for femtosecond lasers, must
  be investigated. Another limitation of the method lies in the computation
  time. For intense lasers, the correction frequency is small (typically a
  fraction of Hertz). \CAMELOT could therefore be used assuming a reasonable
  increase of computation speed. For real-time compensation of atmospheric
  turbulence in a an Adaptive Optics loop, a significant effort is requested
  to manage control frequencies above several hundred of Hertz. Application of
  the method to imaging systems with complicated pupil transmittance is also
  of interest.}

%

%
The authors thank Baptiste Paul for fruitful discussions on classical \PD.
This work has been performed in the framework of the Carnot project SCALPEL. 

\end{document}